\documentclass[a4paper,12pt]{article}
\usepackage{graphicx}
\usepackage{amsmath}
\usepackage{multirow}
\usepackage{float} 
\usepackage{cite}

\begin{document}

\begin{center}
{\large\bf  Discovery potential of the NMSSM CP-odd Higgs at the LHC} \\
\vspace*{1.0truecm}
{\large M. Almarashi\footnote{mmarashi@taibahu.edu.sa}} \\
\vspace*{0.5truecm}
{\it Departement of Physics, Taibah University, Almadinah, KSA}
\end{center}

\vspace*{2.0truecm}
\begin{center}
\begin{abstract}
\noindent 
In this paper we examine the LHC discovery potential of the lightest CP-odd Higgs boson, $a_1$, of the NMSSM produced
in gluon fusion channel $gg\to a_1$. We evaluate the inclusive signal rates of the $a_1$ for a variety of decay channels 
and discuss its possible discovery. 
It is observed that the overall production and decay rates 
at inclusive level are quite sizable and should help extracting the $a_1$ signal over some regions of the NMSSM parameter space.

\end{abstract}
\end{center}

\newcommand{\nn}{\nonumber}
\section{Introduction}
\label{sect:intro}

The observation of a Higgs signal at the LHC \cite{CMS1-Higgs,ATLAS1-Higgs,CMS2-Higgs,ATLAS2-Higgs} in 2012 is
considered a huge success of the Standard Model (SM).
Although its measured properties agrees with the SM predictions, the Higgs-like particle can be embedded in 
Supersymmetric models. In its minimal realisation, the so-called Minimal
Supersymmetric Standard Model (MSSM), the SM is extended by introducing two Higgs
doublets instead of the one in the SM, giving rise to five physical Higgs states: two CP-even Higgses, $h$ and $H$ ($m_h<m_H$), 
one CP-odd Higgs, $A$, and a pair of charged Higgses $H^{\pm}$ instead of the only one in the SM. However, the MSSM suffers
from two serious flaws. The first one is the so-called $\mu$-problem \cite{Kim:1983dt} and the second
one is the little hierarchy problem \cite{Weinberg:1978ym,LlewellynSmith:1981yi}.
The above two problems of the MSSM can be solved by expanding its Higgs sector by introducing an additional Higgs singlet superfield in addition
to the two MSSM-type Higgs doublets. This new model is called the Next-to-Minimal Supersymmetric Standard Model (NMSSM) \cite{review}.
Interestingly, this model can easily accommodate the the measured value of the SM-like Higgs boson mass around 125 GeV
without much fine-tuning
\cite{NMSSMrecent1,NMSSMrecent2,NMSSMrecent3,NMSSMrecent4,
NMSSMrecent5,NMSSMrecent6,NMSSMrecent7,NMSSMrecent8,NMSSMrecent9,NMSSMrecent10,NMSSMrecent11,NMSSMrecent12,NMSSMrecent13,NMSSMrecent14}.

In the NMSSM, the soft Supersymmetry (SUSY)-breaking potential of the Higgs sector is described by the following
Lagrangian contribution
\begin{equation}
V_{\rm NMSSM}=m_{H_u}^2|H_u|^2+m_{H_d}^2|H_d|^2+m_{S}^2|S|^2
             +\left(\lambda A_\lambda S H_u H_d + \frac{1}{3}\kappa A_\kappa S^3 + {\rm h.c.}\right),
\end{equation}
where $H_u$ and $H_d$ are the Higgs doublet fields, $S$ is the singlet one, $\lambda$ and $\kappa$ are Yukawa
coupling parameters while $A_\lambda$ and $A_\kappa$ are dimensionful parameters of the order of SUSY mass scale $M_{\rm{SUSY}}$.

The NMSSM Higgs sector at the tree-level is described by six independent parameters: $\kappa$, $A_\kappa$, $\lambda$, $A_\lambda$, 
tan$\beta$ (the ratio of the vacuum expectation values (VEVs) of the two Higgs doublets) and $\mu_{\rm eff} = \lambda\langle S\rangle$
(where $\langle S\rangle$ is the VEV of the singlet field). Assuming CP-conservation in the Higgs sector, the NMSSM contains seven Higgs states:
three CP-even Higgses $h_{1, 2, 3}$ ($m_{h_1} < m_{h_2} < m_{h_3}$), two CP-odd Higgses $a_{1, 2}$ ($m_{a_1} < m_{a_2} $) and
a pair of charged Higgses $h^{\pm}$. Consequently, the Higgs sector of the NMSSM is phenomenologically richer than that of the MSSM.
Further, when the scalar
component of the singlet field acquires a non-vanishing vacuum expectation value, the $\mu$-term in the superpotential will be
dynamically generated, thus solving the $\mu$-problem \cite{Ellis:1988er}.

In the NMSSM, $a_1$ state is a composition of the usual doublet component of the CP-odd MSSM Higgs boson, $a_{\rm MSSM}$, and
the new singlet component, $a_{\rm S}$, coming from the singlet superfield of the NMSSM. This can be written as \cite{excess}:
\begin{equation}
 a_1=a_{\rm MSSM}\cos\theta_A+a_{\rm S}\sin\theta_A,
\end{equation}
where $\cos\theta_A$ and $\sin\theta_A$ are the mixing angles. When $\cos\theta_A$ goes to zero, the $a_1$ is highly singlet. 
To a good approximation the $m_{a_1}$ can be written as:
\begin{equation} 
 m^2_{a_1}=\frac{9A_\lambda \mu_{\rm eff}}{2\sin2\beta}\cos^2\theta_A-3\frac{\kappa A_\kappa \mu_{\rm eff}}{\lambda}\sin^2\theta_A.
 \end{equation}
It is clear from the last equation that all the tree level parameters of the NMSSM Higgs sector jointly affects $m_{a_1}$ in general.

There have been some studies 
exploring the production potential of the NMSSM CP-odd Higgs states at the LHC and other colliders
\cite{Hiller:2004ii,Arhrib:2006sx,Domingo:2008rr,Almarashi:2010jm,Almarashi:2011bf,Almarashi:2011hj,Almarashi:2011te,Almarashi:2011qq,NMSSMreviewed3,
Cerdeno:2013qta,King:2014xwa,Bomark:2014gya,Ellwanger:2016qax,Conte:2016zjp,Guchait:2016pes}.  It was found that the best direct production
channel of producing the lightest CP-odd Higgs state $a_1$ is through its production in association with a bottom – antibottom pair $b \bar ba_1$.
However, there is still no a comprehensive study about the detectability of the CP-odd Higgs particles at the LHC in gluon fusion production channel 
$gg \to a$.

In this paper, we explore the discovery potential of the $a_1$  at the LHC by looking for its direct production rather than looking
for its traditional production through $h_{1, 2}$ decay. We  examine the discovery potential of the $a_1$  produced
in gluon fusion channel $gg \to a_1$ through a variety of decay modes. 
We will estimate the inclusive cross section of the $a_1$ production to examine
whether or not there exist some parameter space where the $a_1$ can be discovered at the LHC although detail analysis of the background is
required to make a conclusive decision. It is shown that there is some regions of the NMSSM parameter space where the cross sections times branching
ratio rates are quite remarkable for some decay channels and can be useful to look for the $a_1$ signals at the LHC through the above production
channel.
  
The plan of this paper is as follows. In section ~\ref{sect:scan} we describe the parameter space scans
performed in the context of the NMSSM and discuss the allowed decay channels of the $a_1$. In section ~\ref{sect:rates} we present the inclusive event rates of $a_1$ production
at the LHC for various decay channels.
Finally, we summarize our results in section ~\ref{sect:summa}.

\section{\large Exploring the NMSSM Parameter Space}
\label{sect:scan}

For our study of the NMSSM Higgs sector
we have used the package NMSSMTools5.1.2 \cite{NMHDECAY1,NMHDECAY2,NMSSMTools} for the calculation of the branching ratios and spectrum of Higgs bosons
and SUSY particles of the NMSSM. The package systematically takes into account various theoretical and experimental constraints.

We have used the above package to scan over some regions of the NMSSM parameter space in order to obtain a general view of the  phenomenology of
the lightest CP-odd Higgs boson, $a_1$, at the LHC. We have set the six tree level parameters in the following ranges:

\begin{center}
$0.001 \leq \lambda \leq  0.7$, \phantom{aa} $-0.65 \leq \kappa \leq  0.65$,\phantom{aa} $1.6 \leq \tan\beta \leq  60$, \phantom{aa} \\
$100 \leq \mu \leq  1000$ GeV, \phantom{aa} $-2000 \leq A_{\lambda} \leq  2000$ GeV,\phantom{aa} $-20 \leq A_{\kappa} \leq  20$ GeV. \\
\end{center}
Notice that we have chosen the small values of $A_{\kappa}$ to obtain small values of the $m_{a_1}$. 
Remaining soft SUSY breaking right- and left-handed masses for the first two generations and for the third generation, 
soft SUSY breaking trilinear couplings and gaugino soft SUSY breaking masses,
contributing
at higher order level, have been set as:\\
$\bullet\phantom{a}m_Q = m_U = m_D = m_L = m_E = 1$ TeV,\\
$\bullet\phantom{a}m_{Q_3} = m_{U_3} = m_{D_3} = m_{L_3} = m_{E_3} = 1$ TeV, \\
$\bullet\phantom{a}A_{U_3} = A_{D_3} = A_{E_3} = -2$ TeV,\\
$\bullet\phantom{a} M_1 = 150$ GeV, $M_2 = 300$ GeV,  $M_3 = 1$ TeV.\\

The interesting decay channels which may help to discover the lightest CP-odd neutral Higgs boson $a_1$ at the LHC are:
\begin{equation}
a_1\rightarrow \mu^+\mu^-,\tau^+\tau^-,gg,
s\bar s,c\bar c,b\bar b, t\bar t, 
\gamma\gamma,Z\gamma,~{\rm sparticles}.
\end{equation}

We have performed a random scan  over one million  points in the specified parameter space. The scan output contains masses, branching ratios 
and couplings of the NMSSM Higgses and SUSY particles for all the surviving points which have passed 
the various experimental and theoretical constraints.

\section{\large Higgs boson Signal Rates}
\label{sect:rates}

In order to investigate the discovery potential of the $a_1$ at the LHC, we have computed 
the inclusive production rates by multiplying the NMSSM gluon fusion
production cross section using CalcHEP \cite{CalcHEP}  with the branching ratios computed with
the NMSSMTools for all surviving data points \footnote{We assume  a centre-of-mass 
energy $\sqrt s=14$ TeV for the LHC.}.

Figure 1 shows the production rates in femtobarn (fb) for the $a_1$ in the $\tau^+\tau^-$ and $\mu^+\mu^-$ 
final states, $\sigma(gg\to a_1)~{\rm Br}(a_1\to \tau^+\tau^-)$ and
$\sigma(gg\to a_1)~{\rm Br}(a_1\to \mu^+\mu^-)$, as functions of $m_{a_1}$ and of the corresponding branching ratios.
As expected, the signal rate decreases with increasing $m_{a_1}$, see the left-panels of the figure. 
Also, it is remarkable to notice that the signal rates into $\tau^+\tau^-$ and $\mu^+\mu^-$ are sizable, topping the
$7.3\times 10^5$ fb and $2.8\times 10^3$ fb levels for $\tau^+\tau^-$ and $\mu^+\mu^-$ final states, respectively,
for small values of $m_{a_1}$ and decreasing 
rapidly with increasing $m_{a_1}$ (left-panels). Further, it is clear that
the rate into $\mu^+\mu^-$ final state shows the same pattern as the ones into $\tau^+\tau^-$ but is suppressed by 
a factor of $\approx$ $(m_\mu^{\rm pole} / m_\tau^{\rm pole})^2$. 
Also, notice that Br $(a_1\to \tau^+\tau^-)$ and Br $(a_1\to \mu^+\mu^-)$
reaches about 10\% and 0.035\%, respectively, in most of the parameter space that has $m_{a_1}\geq 10$ GeV, in which case the $a_1$ decay into 
$b\bar b$ is kinematically open and dominant. However, there is a few points with light $a_1$, $m_{a_1}< 10 $ GeV,
yielding large Br$(a_1\to \tau^+\tau^-)$  and Br$(a_1\to \mu^+\mu^-)$ greater than $90\%$ and $0.4\%$, respectively, 
see the right-panels of the figure \footnote{Notice that the mass region below the $b\bar b$ threshold is severely 
constrained, see, e.g., Ref. \cite{Lebedev}
(and references therein).}. 

Overall, the inclusive cross section $\sigma(gg\to a_1)~{\rm Br}(a_1\to \tau^+\tau^-)$ is quite large,
so  the $\tau^+\tau^-$ decay channel could be a good channel to discover 
the $a_1$ at the LHC, assuming the double- or single-leptonic decay channels of the $\tau$'s. As for the $\mu^+\mu^-$ final state,
the rates $\sigma(gg\to a_1)~{\rm Br}(a_1\to \mu^+\mu^-)$ is also remarkable for $a_1$ low mass.
For example, for $m_{a1}$ $\sim$ 100 GeV,
one can have up to several thousands events in the $\mu^+\mu^-$ channel at 100 $fb^{-1}$ integrated luminosity. Therefore, 
the $\mu^+\mu^-$ decay channel is a promising channel to discover a light $a_1$.

The signal rates for $a_1$ decaying into $b\bar b$ and $t\bar t$ as functions of $m_{a_1}$ and of the corresponding branching ratios are
shown in figure 2. The top-left panel of the figure shows that the signal rate in the $b\bar b$ final state is quite large, as the rates
are at nb level for $m_{a1} \leq 100$ GeV and at fb level for $m_{a1} \simeq 300$ GeV. Also, it is clear from the top-right panel of the figure 
 that for most points of the NMSSM parameter space
the $a_1$ dominantly decays into $b\bar b$ final states with
branching fraction close to 90\%. Although, the discovery of the $a_1$ through $b\bar b$ channel is challenging
due to large backgrounds, the size of the inclusive cross section is large enough to discover the $a_1$ especially if the background
can be successfully reduced to manageable levels. As for the top quark pair final states, the signal rates are small, below 1 fb level, see the bottom-panels of the figure.
 Therefore, we think that the $a_1$  production at the LHC through $t\bar t$ final state is challenging
 due to the smallness of the signal rate and its complicated final state \footnote{We do 
not study the inclusive production rates for the decays
$a_1\rightarrow s\bar s$, $a_1\rightarrow 
c\bar c$ and $a_1\rightarrow gg$  due to large QCD backgrounds and smallness of their production rates in general}.

Figure 3 shows the distribution of event rates $\sigma(gg\to a_1)~{\rm Br}(a_1\to \gamma\gamma)$ and $\sigma(gg\to a_1)~{\rm Br}(a_1\to Z\gamma)$
as functions of $m_{a_1}$ and of the corresponding branching ratios ${\rm Br}(a_1\to \gamma\gamma)$ and of $~{\rm Br}(a_1\to Z\gamma)$.
It is remarkable to
notice that the rates are quite large, reaching maximum rates of about several hundreds fb for $\gamma\gamma$ and several tens fb
for $Z\gamma$ final states. This is quite interesting as such signal events may be detectable at planned LHC
luminosities. Also, it is obvious that the Br$(a_1\to \gamma\gamma)$ and 
Br$(a_1\to Z\gamma)$ can be dominant in some regions of the NMSSM parameter space, see the right-panes of the figure. 
This occurs when the $a_1$ is mostly singlet-like with a very small doublet component, i.e. the mixing angle $\cos\theta_A$ is very small, see
figure 4. Such a possibility emerges when the tree-level decay to fermion-antifermion are highly suppressed, and so 
the decays $a_1\to \gamma\gamma$ and $a_1\to Z\gamma$ 
are dominant, due to large contributions from chargino loops.

Higgs bosons decaying into supersymmetric particles could play important roles for searching for such bosons. For example, 
to study the possibility of the $a_1$ production to SUSY particles, we have calculated the inclusive signal rates 
for the $a_1$ decaying into the lightest neuralinos $\chi^0_1 \chi^0_1$ 
and into charginos $\chi^+_1 \chi^-_1$, see figure 5.  It is shown that the maximum signal rates are a few tens for the former and a few fb for the latter.
The right-panels of the figure show that the ${\rm Br}(a_1\to \chi^0_1 \chi^0_1)$ and ${\rm Br}(a_1\to \chi^+_1 \chi^-_1 )$ can be dominant
in large regions of the NMSSM parameter space due to the enhancement of $a_1$ couplings with singlino-Higgsino components
in $\chi^0_1$, $\chi^+_1 $ and $\chi^-_1$ states \footnote{The $a_1$ decay into lightest neutralino contributes to
its invisible decay width.}. Notice that if the $a_1$ is highly singlet and its decays into $\chi^0_1 \chi^0_1$ or into $\chi^+_1 \chi^-_1$ are
kinematically open, the ${\rm Br}(a_1\to \gamma\gamma)$ and ${\rm Br}(a_1\to Z\gamma)$ are
no longer dominant, see figure 6, where the maximum values of both
the ${\rm Br}(a_1\to \gamma\gamma)$ and ${\rm Br}(a_1\to Z\gamma)$ are $\leq$ 1\% (top-panels) and $\leq$ 0.01\% (bottom-panels),
in which case the $a_1$ decay into $\chi^0_1 \chi^0_1$ or into $a_1\to \chi^+_1 \chi^-_1$ becomes dominant. 

Assuming R-parity is conserved,  the lightest Supersymmetric particle (LSP) could be
the lightest neutralino $\chi^0_1$ in large regions of the parameter space of the NMSSM. If this particle is highly singlet, it will
easily reach Dark Matter (DM) relic density and become an ideal particle for cold dark matter. So, the generic signatures of 
supersymmetric particles will involve missing energy due to the two neutral stable $\chi^0_1$s
which escape the  detector making the discovery of the $a_1$ at the LHC is challenging.

\section{Conclusions}
\label{sect:summa}
The SM-like Higgs boson mass of the range $\sim $ 125 GeV can be accommodated in the framework of the NMSSM without much fine-tuning. In
this model, by assuming CP-conservation, there are seven Higgs bosons: three CP-even, two CP-odd and a pair of charged Higgses. We have
investigated the discovery potential of the lightest CP-odd Higgs boson $a_1$  produced
through the gluon fusion production channel $gg \to a_1$  at the LHC with high center of mass energy, $\sqrt{s} = 14$ TeV in the context of the NMSSM.
After computing the inclusive signal rates for  $\tau^+\tau^-$, $\mu^+\mu^-$, $b \bar b$, $t \bar t$, $\gamma\gamma$, $Z\gamma$,
$\chi^0_1 \chi^0_1$ and $\chi^+_1 \chi^-_1$ decay channels, we have found that the $a_1$ can have sizable signal
rates in some regions of the NMSSM parameter space. While further studies about signal-to-background analysis are needed to make a final
decision, we believe that the promising decay channels for the $a_1$ discovery at the LHC in the gluon fusion channel are 
$\tau^+\tau^-$, $\mu^+\mu^-$, $b \bar b$, $\gamma\gamma$, $Z\gamma$ channels.

Furthermore, we have noticed that the Br$(a_1\to \gamma\gamma)$, Br$(a_1\to Z\gamma)$, 
Br$(a_1\to \chi^0_1 \chi^0_1)$ and Br$(a_1\to \chi^+_1 \chi^-_1)$
can be dominant in some regions of the NMSSM parameter space.  However, the Br$(a_1\to \gamma\gamma)$ and Br$(a_1\to Z\gamma)$
can only be dominant if the $a_1$ is highly singlet with a very small 
doublet component and both the channels
$a_1\to \chi^0_1 \chi^0_1$ and $a_1\to \chi^+_1 \chi^-_1$ are kinematically closed.
If the $a_1$ is a singlet-like and the channel $a_1\to \chi^0_1 \chi^0_1$ or $a_1\to \chi^+_1 \chi^-_1$ is kinematically open, 
the $a_1$ decays into $\gamma\gamma$ and into $ Z\gamma$ are no
longer dominant, in which case the $a_1$ decay into $\chi^0_1 \chi^0_1$ or into $a_1\to \chi^+_1 \chi^-_1$ becomes substantial and dominant.

\section*{Acknowledgments}
This work is funded by Taibah University, KSA.

\begin{figure}
 \centering\begin{tabular}{cc}
  \includegraphics[scale=0.60]{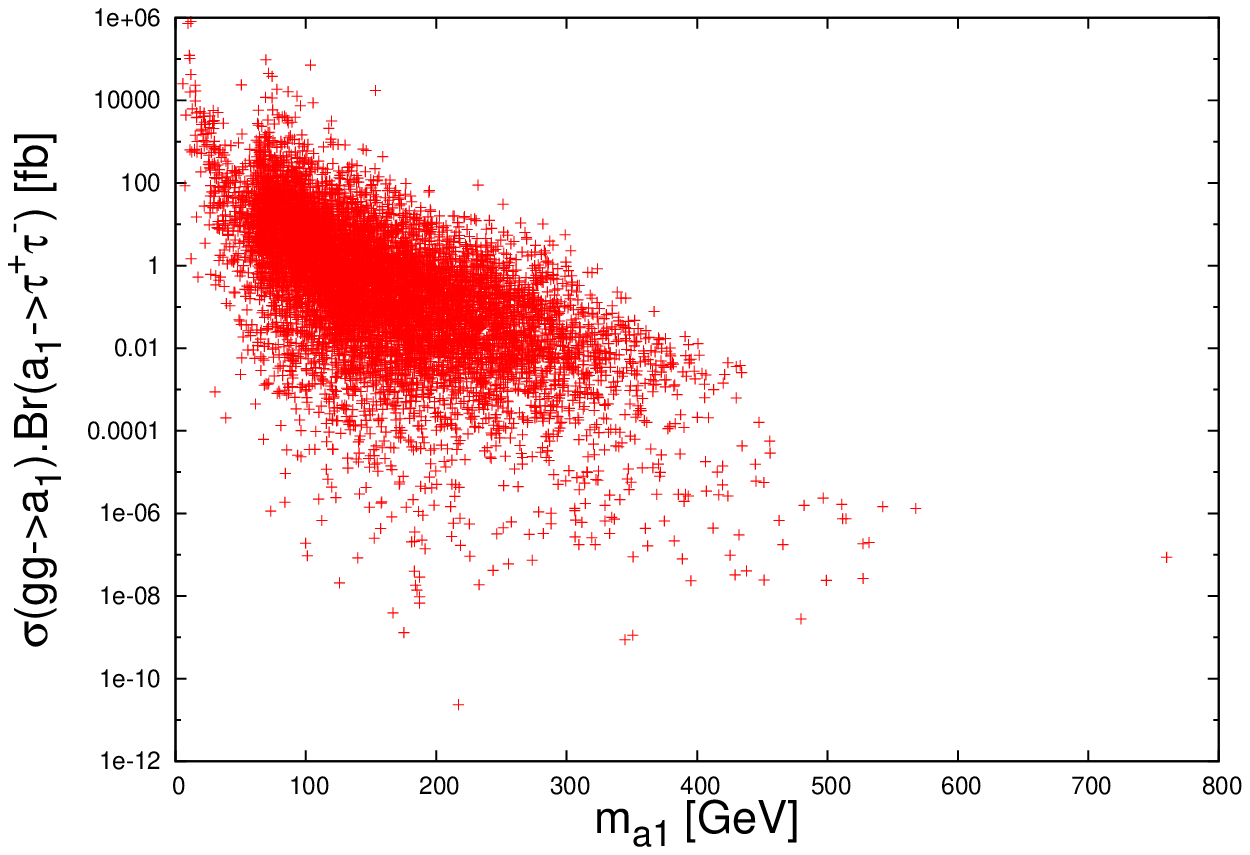}&\includegraphics[scale=0.60]{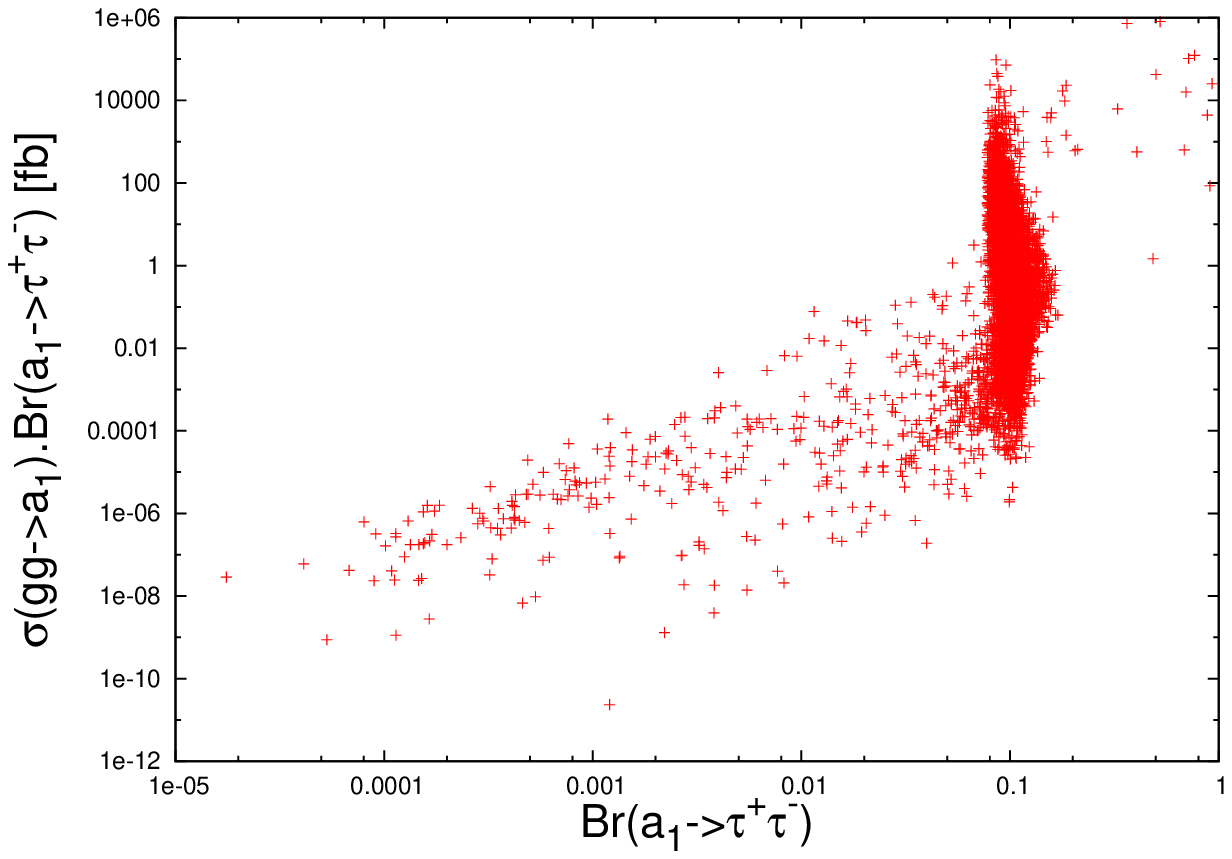}\\
  \includegraphics[scale=0.60]{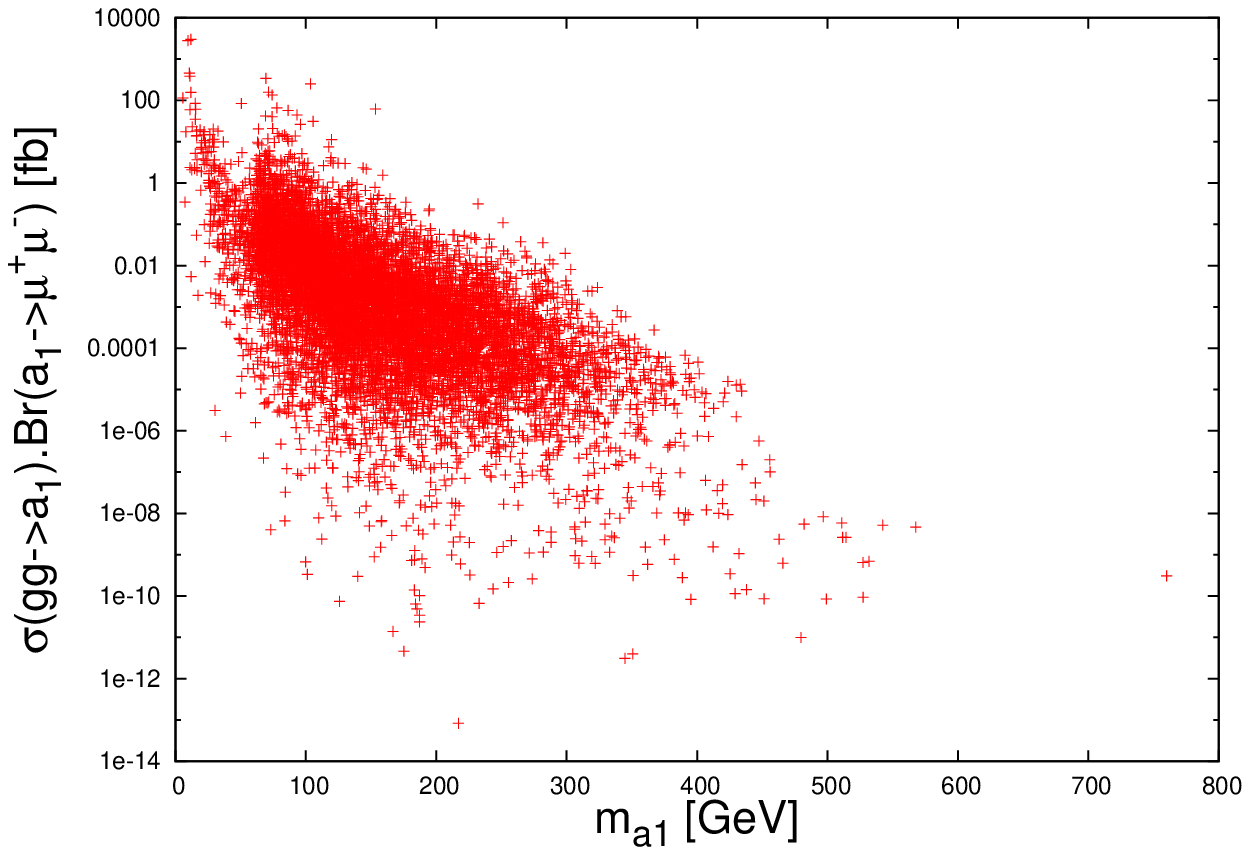}&\includegraphics[scale=0.60]{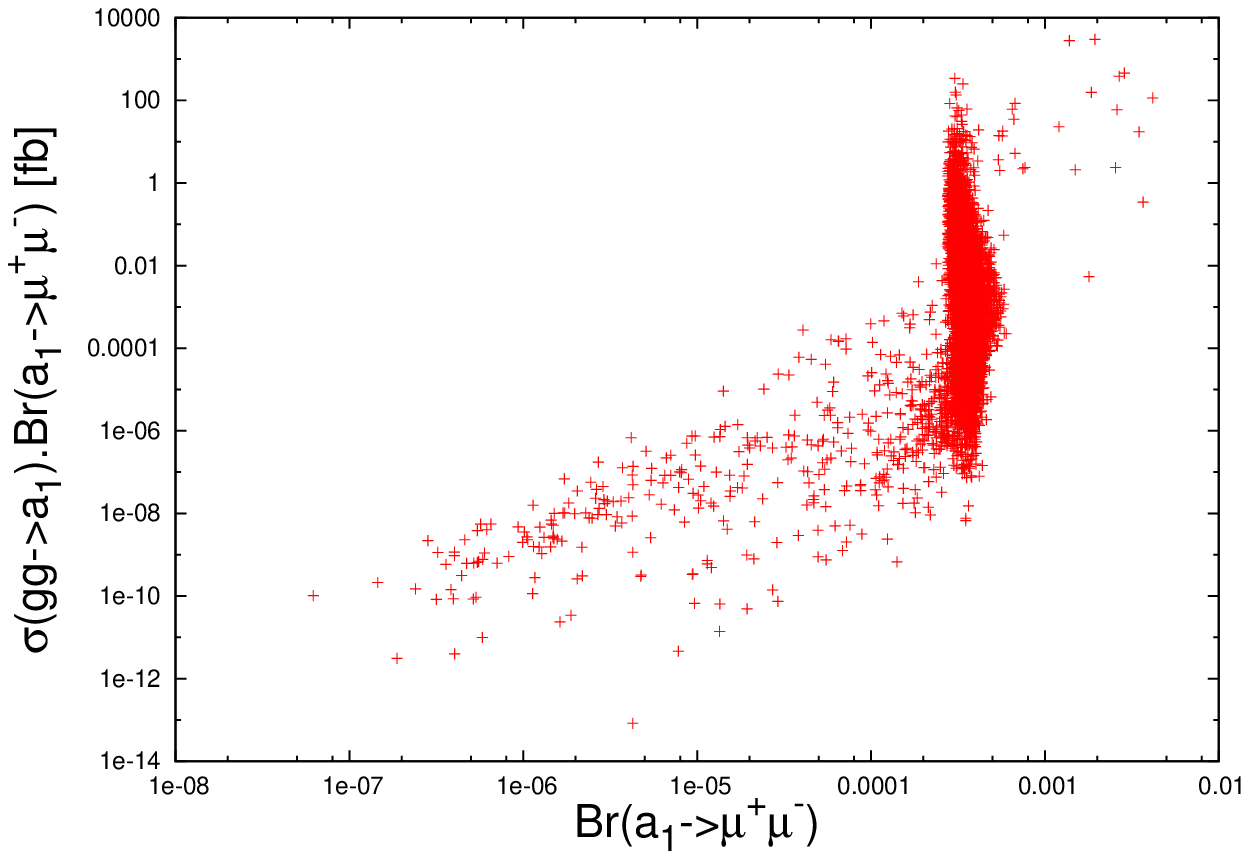}
    
 \end{tabular}
\label{fig:sigma-scan1}
 
\caption{The production rates for $\sigma(gg\to a_1)~{\rm Br}(a_1\to \tau^+\tau^-)$ (top)
and $\sigma(gg\to a_1)~{\rm Br}(a_1\to \mu^+\mu^-)$ (bottom) as functions 
of $m_{a_1}$ (left) and of corresponding branching fractions (right).}
\end{figure}

\begin{figure}
 \centering\begin{tabular}{cc}
  \includegraphics[scale=0.60]{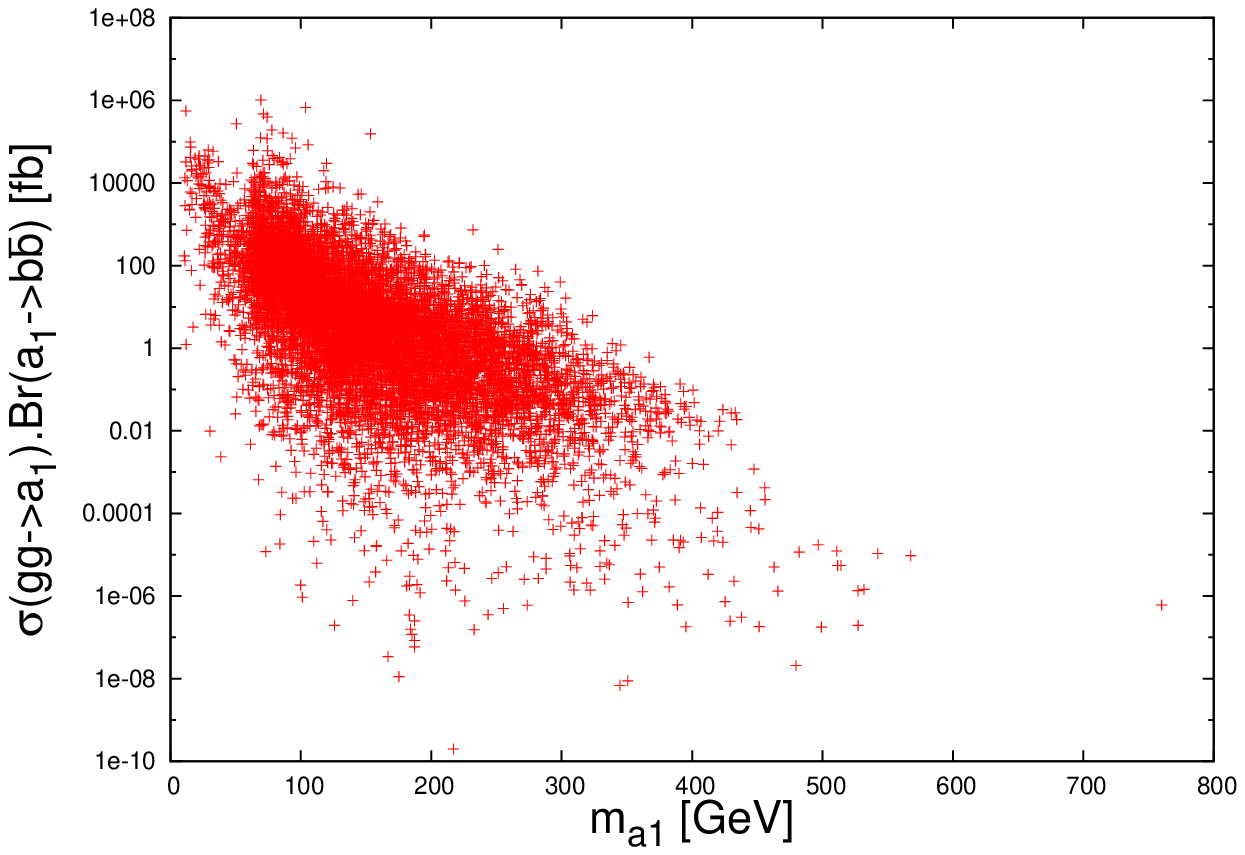}&\includegraphics[scale=0.60]{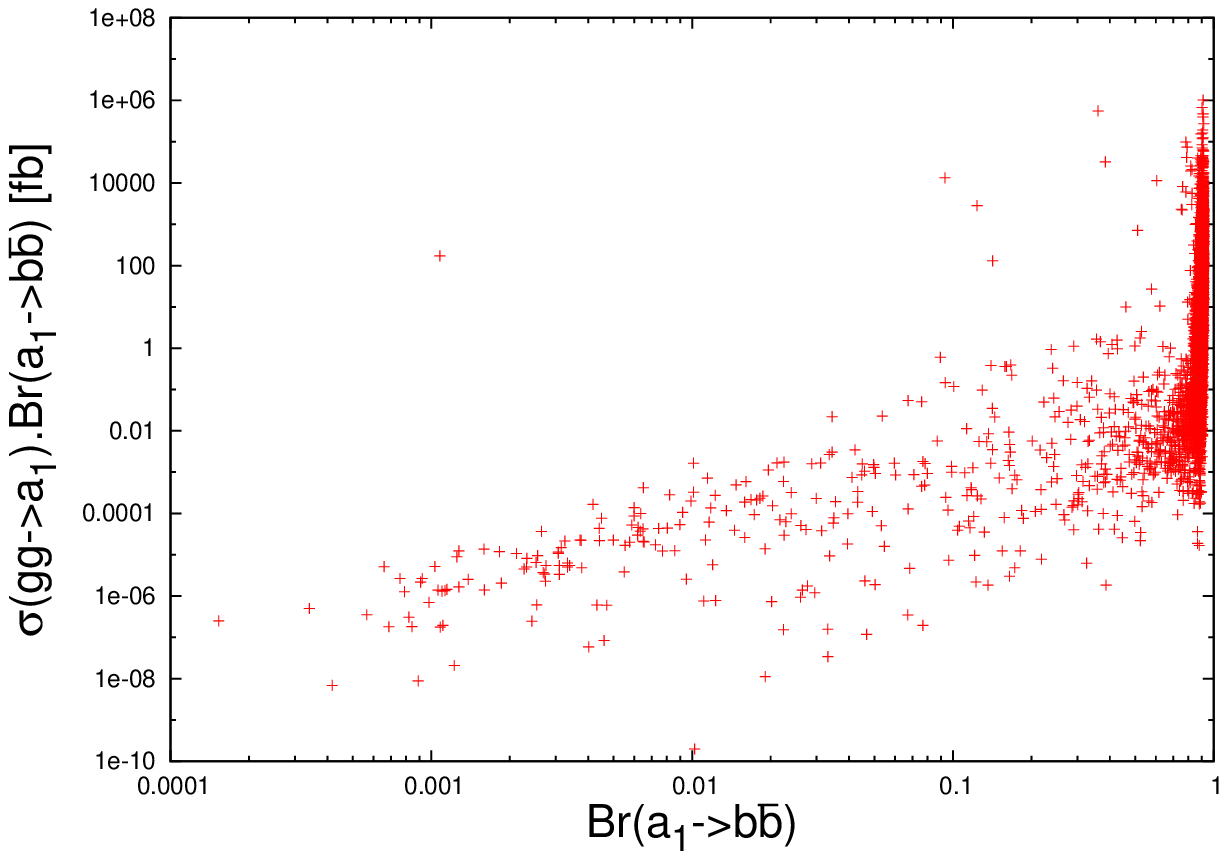}\\
  \includegraphics[scale=0.60]{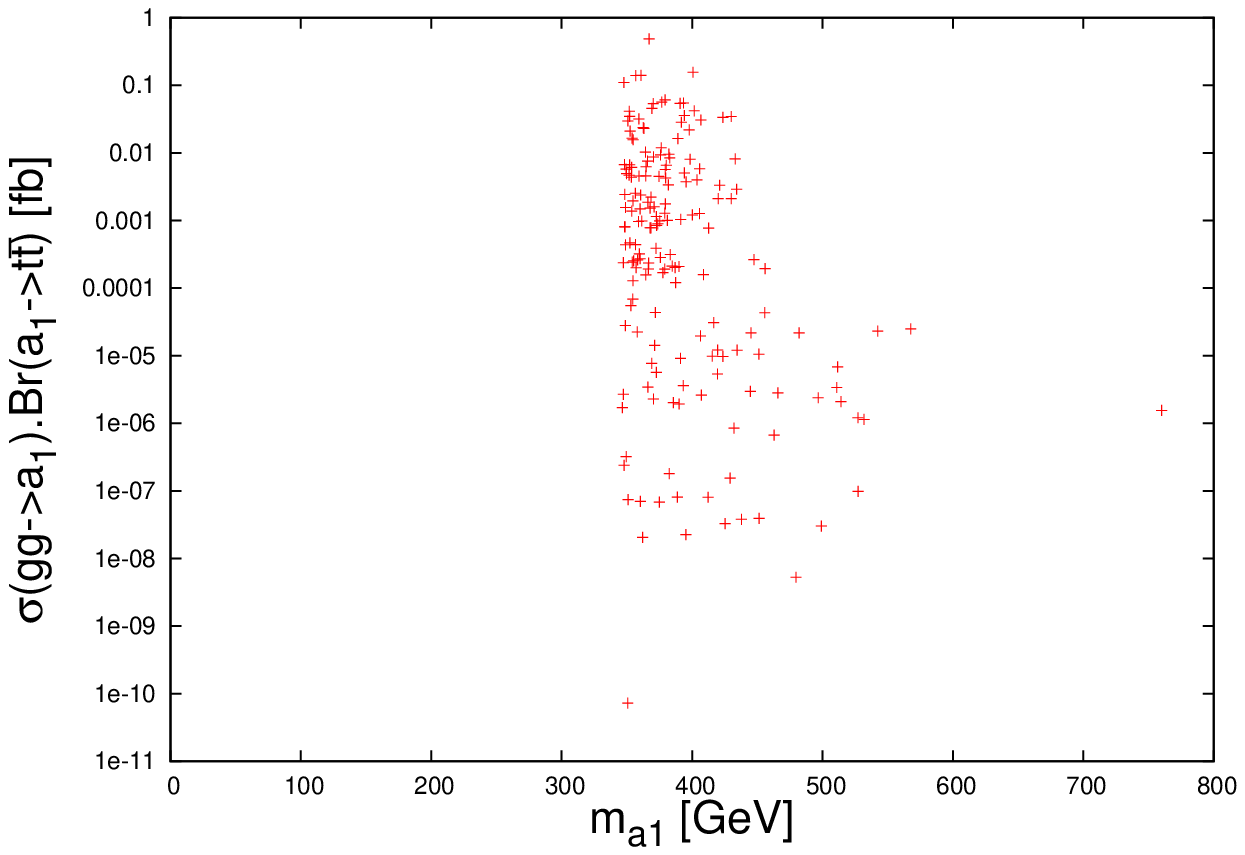}&\includegraphics[scale=0.60]{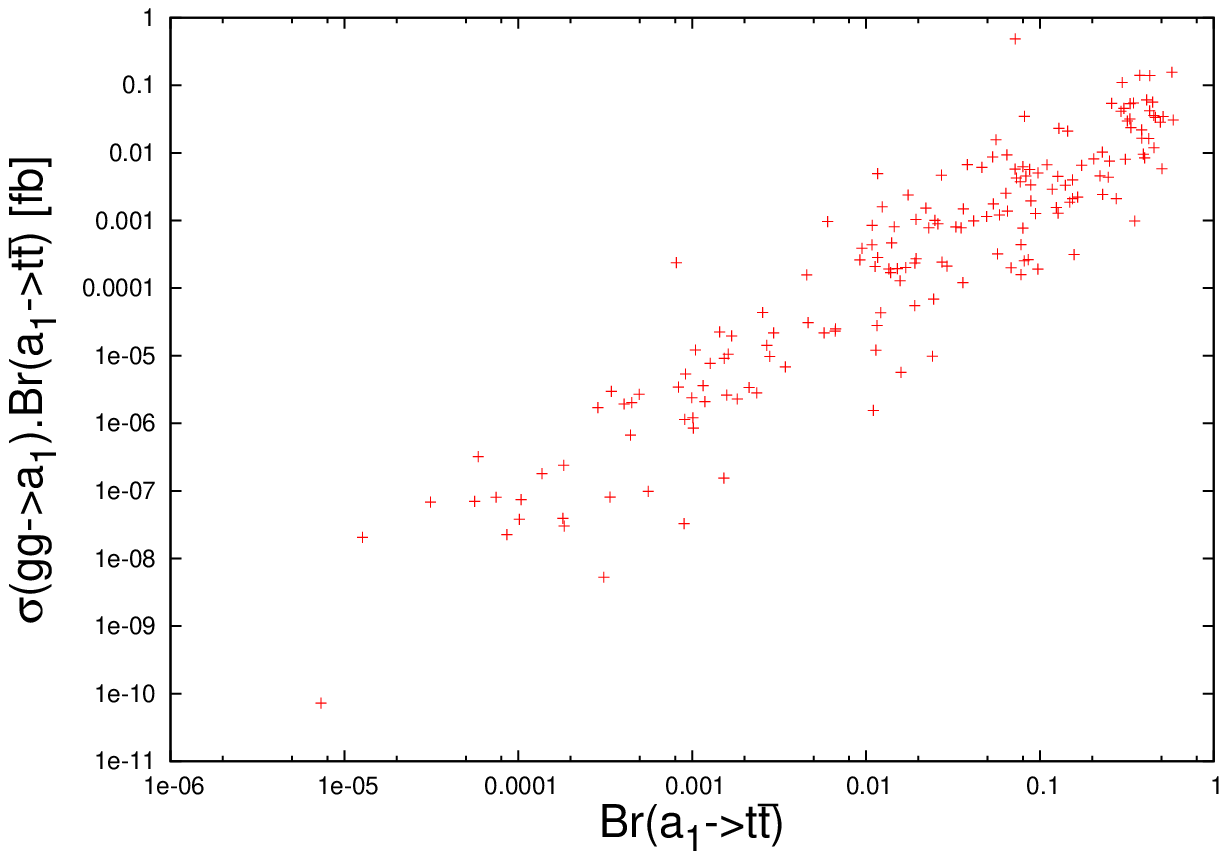}
    
 \end{tabular}
\label{fig:sigma-scan2}
\caption{The production rates for $\sigma(gg\to a_1)~{\rm Br}(a_1\to b\bar b)$ (top)
and $\sigma(gg\to a_1)~{\rm Br}(a_1\to t\bar t)$ (bottom) as functions 
of $m_{a_1}$ (left) and of corresponding branching fractions (right).}

\end{figure}

\begin{figure}
 \centering\begin{tabular}{cc}
  \includegraphics[scale=0.60]{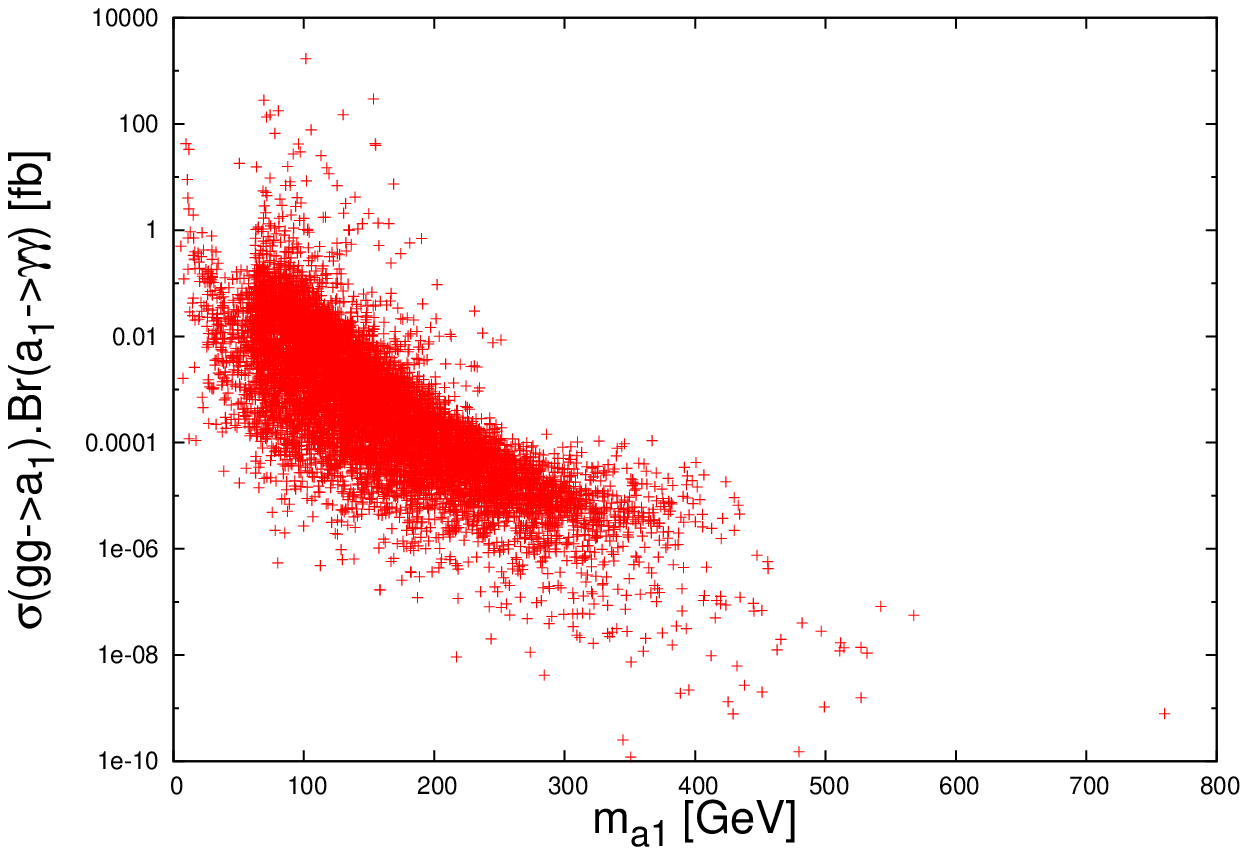}&\includegraphics[scale=0.60]{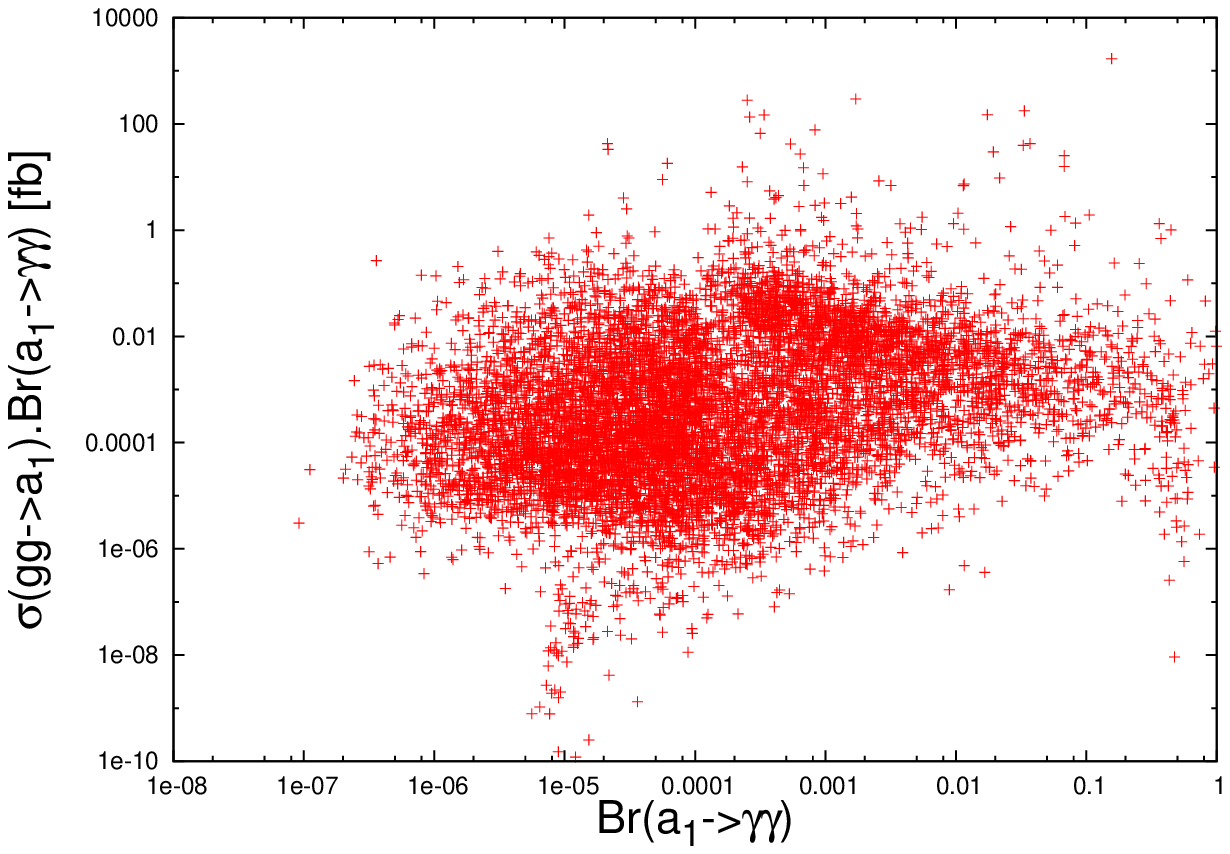}\\
  \includegraphics[scale=0.60]{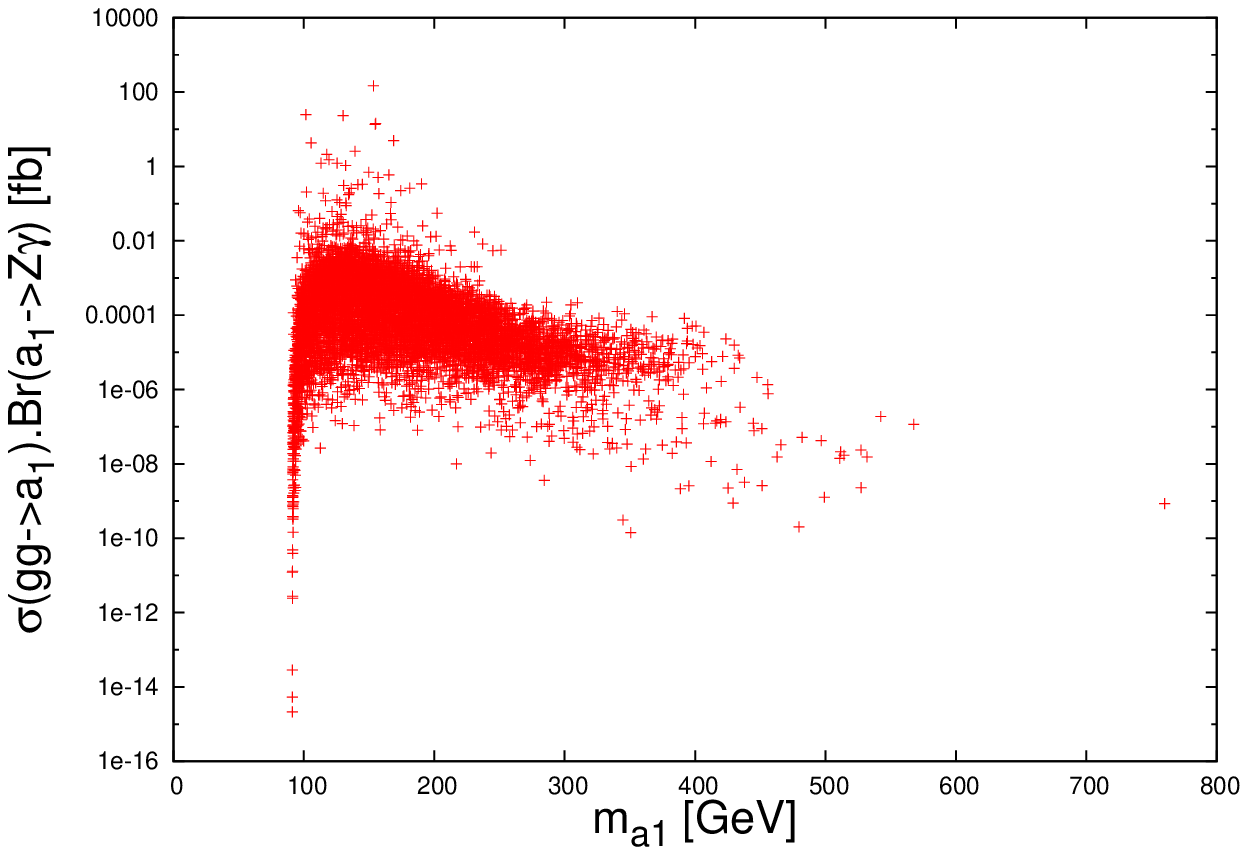}&\includegraphics[scale=0.60]{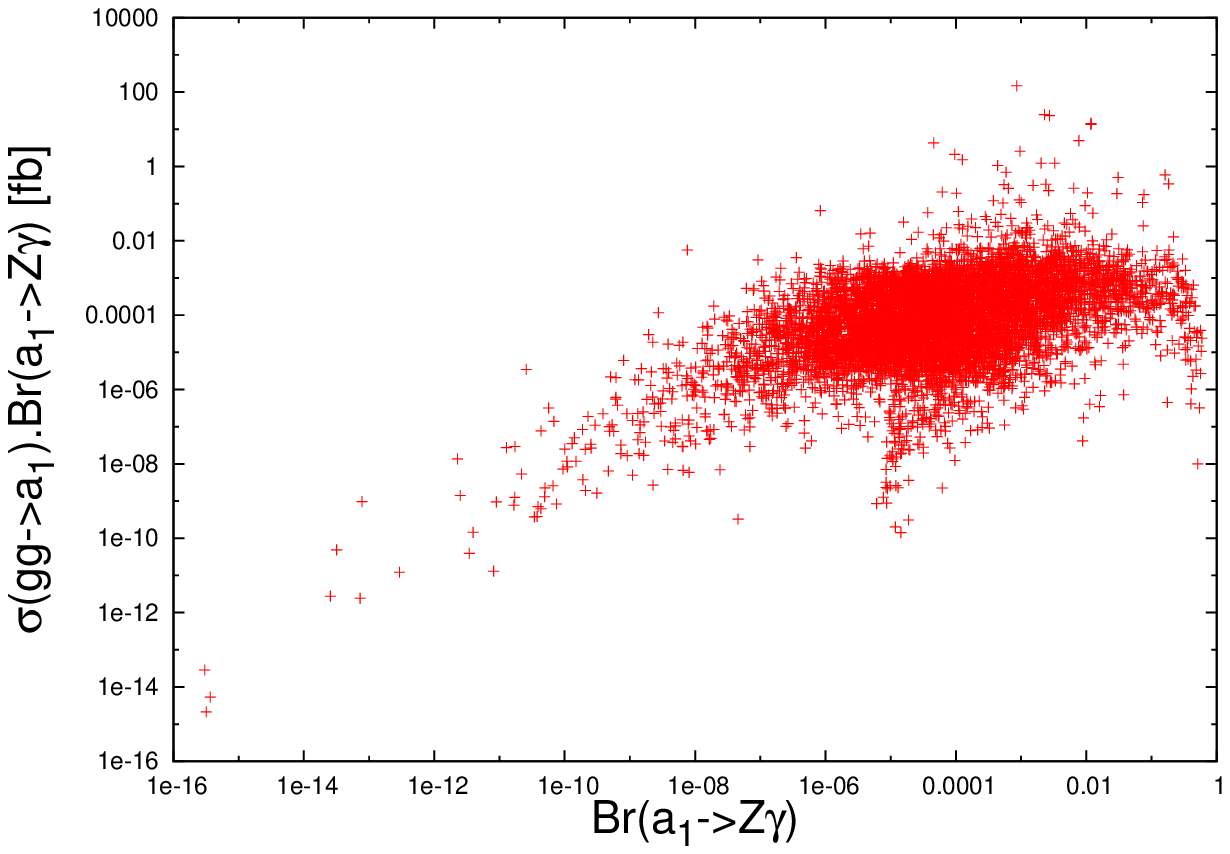}

 \end{tabular}
\label{fig:sigma-scan3}
\caption{The production rates for $\sigma(gg\to a_1)~{\rm Br}(a_1\to \gamma\gamma)$ (top) and 
$\sigma(gg\to a_1)~{\rm Br}(a_1\to Z\gamma)$ (bottom) as functions 
of $m_{a_1}$ (left) and of corresponding branching fractions (right). }
\end{figure}

\begin{figure}
 \centering\begin{tabular}{cc}
  \includegraphics[scale=0.60]{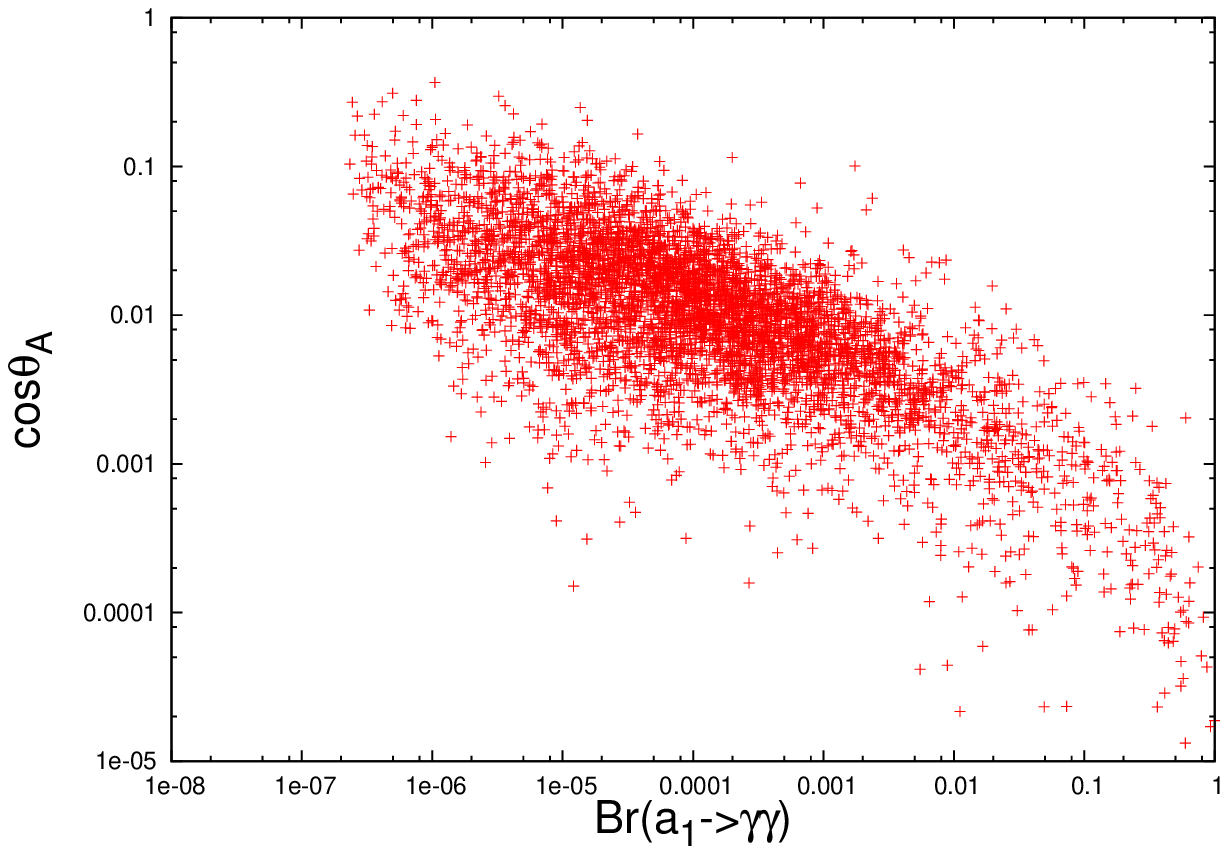}&\includegraphics[scale=0.60]{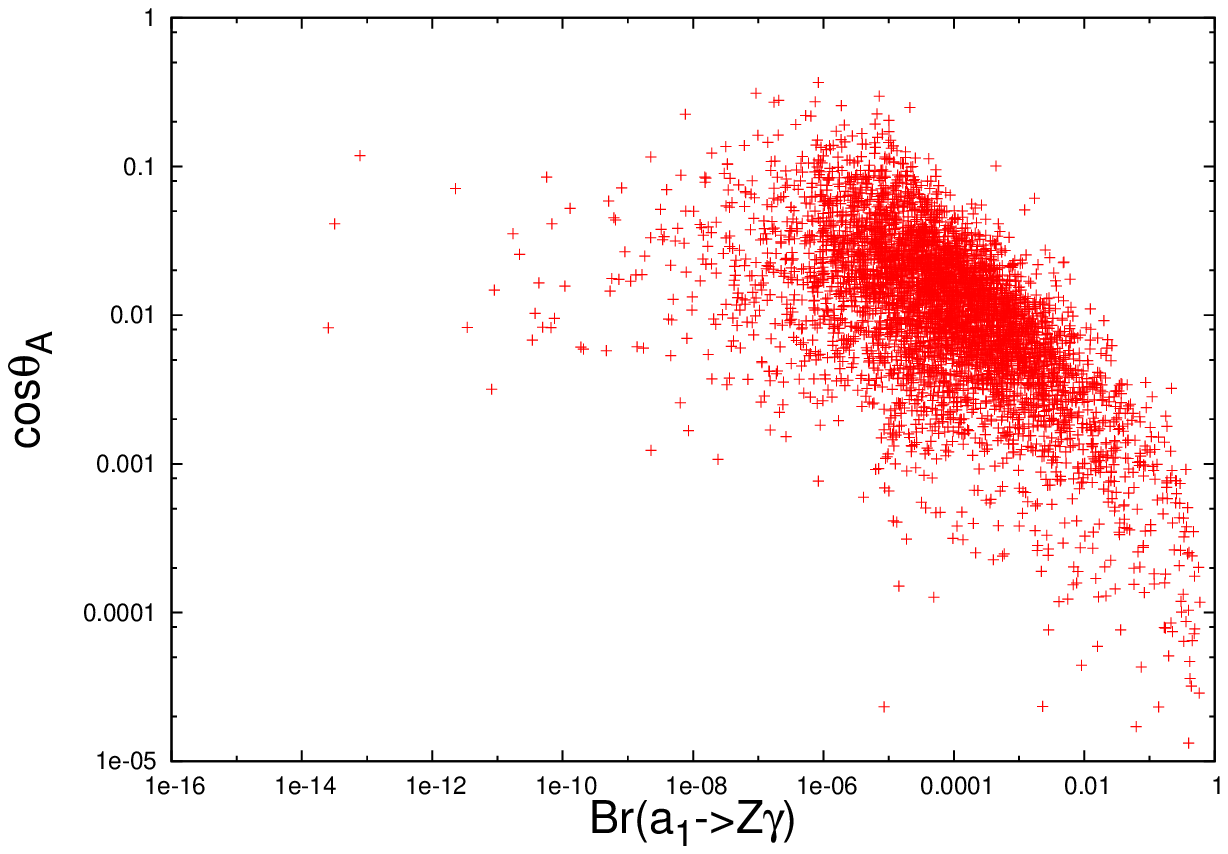}  
 \end{tabular}
\label{fig:sigma-scan4}
\caption{ The correlations between
the mixing angle $\cos\theta_A$ and the branching fractions ${~\rm Br}(a_1\to \gamma\gamma)$ (left) 
and ${~\rm Br}(a_1\to Z\gamma)$ (right). }
\end{figure}

\begin{figure}
 \centering\begin{tabular}{cc}
  \includegraphics[scale=0.60]{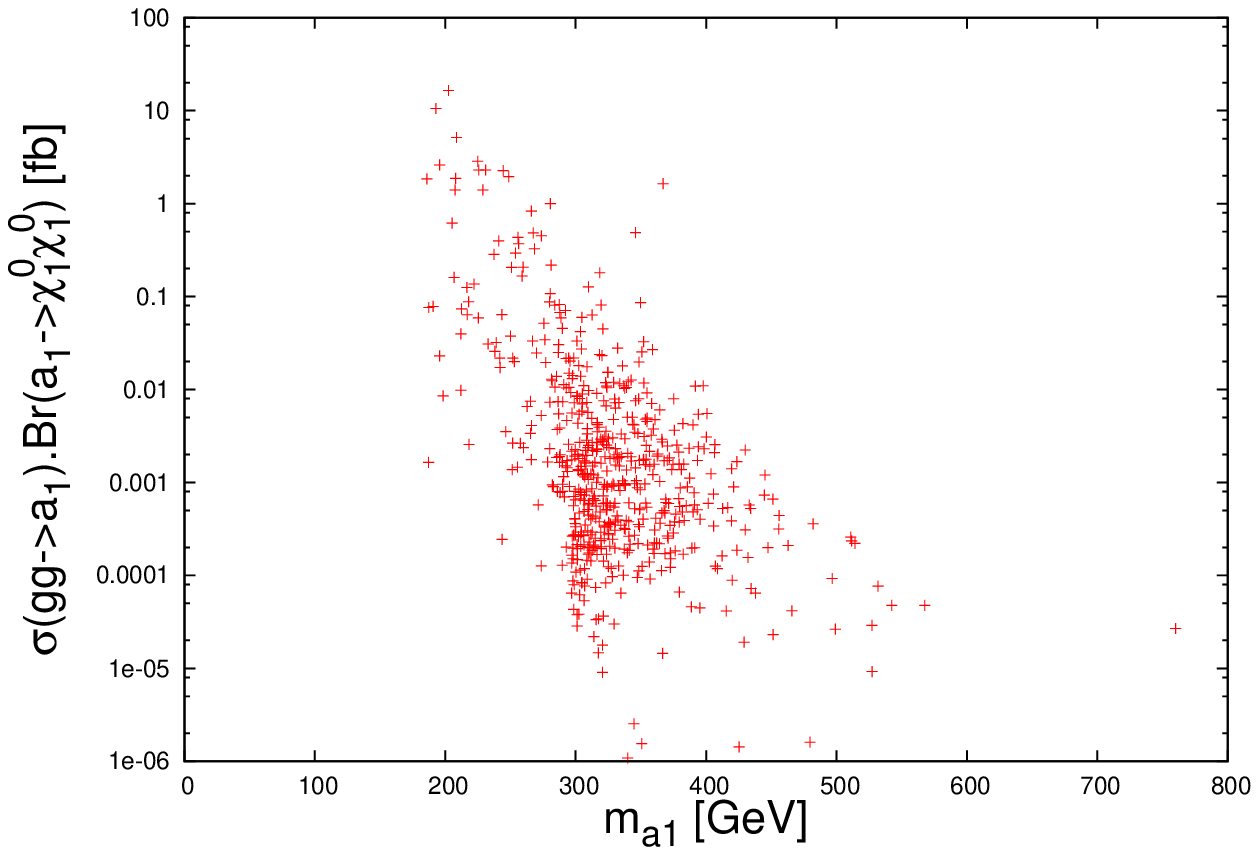}&\includegraphics[scale=0.60]{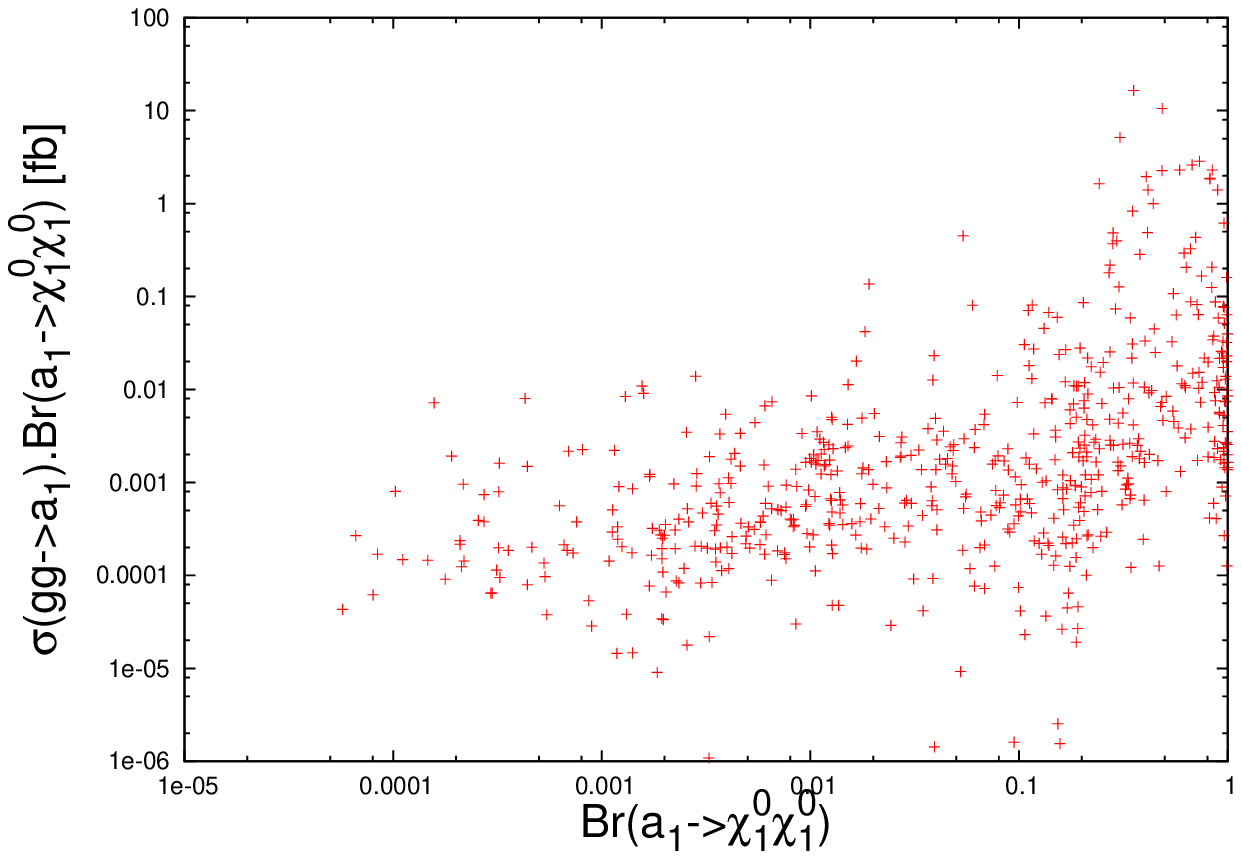}\\
  \includegraphics[scale=0.60]{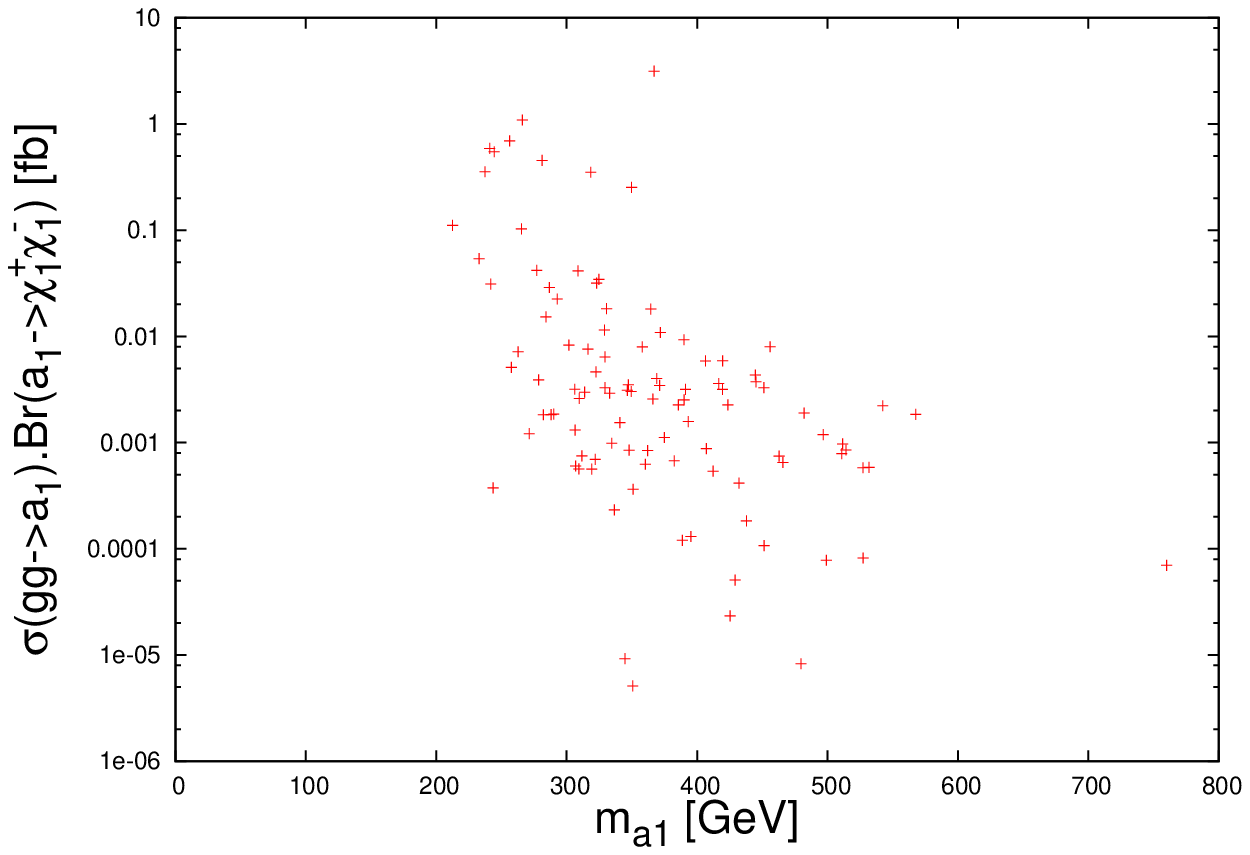}&\includegraphics[scale=0.60]{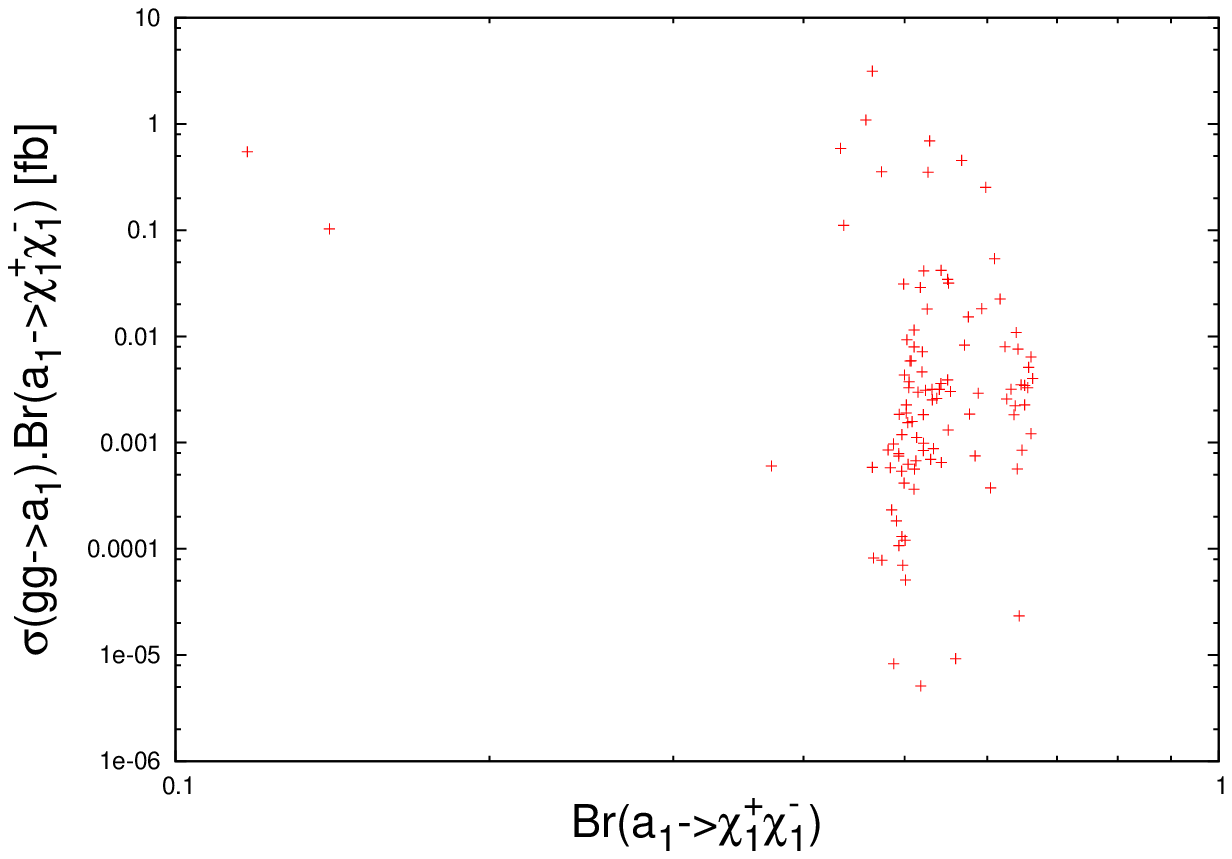}

 \end{tabular}
\label{fig:sigma-scan4}
\caption{The production rates for $\sigma(gg\to a_1)~{\rm Br}(a_1\to \chi^0_1 \chi^0_1)$ (top)
and $\sigma(gg\to a_1)~{\rm Br}(a_1\to \chi^+_1 \chi^-_1 )$ (bottom) as functions 
of $m_{a_1}$ (left) and of corresponding branching fractions (right).}

\end{figure}

\begin{figure}
 \centering\begin{tabular}{cc}
  \includegraphics[scale=0.60]{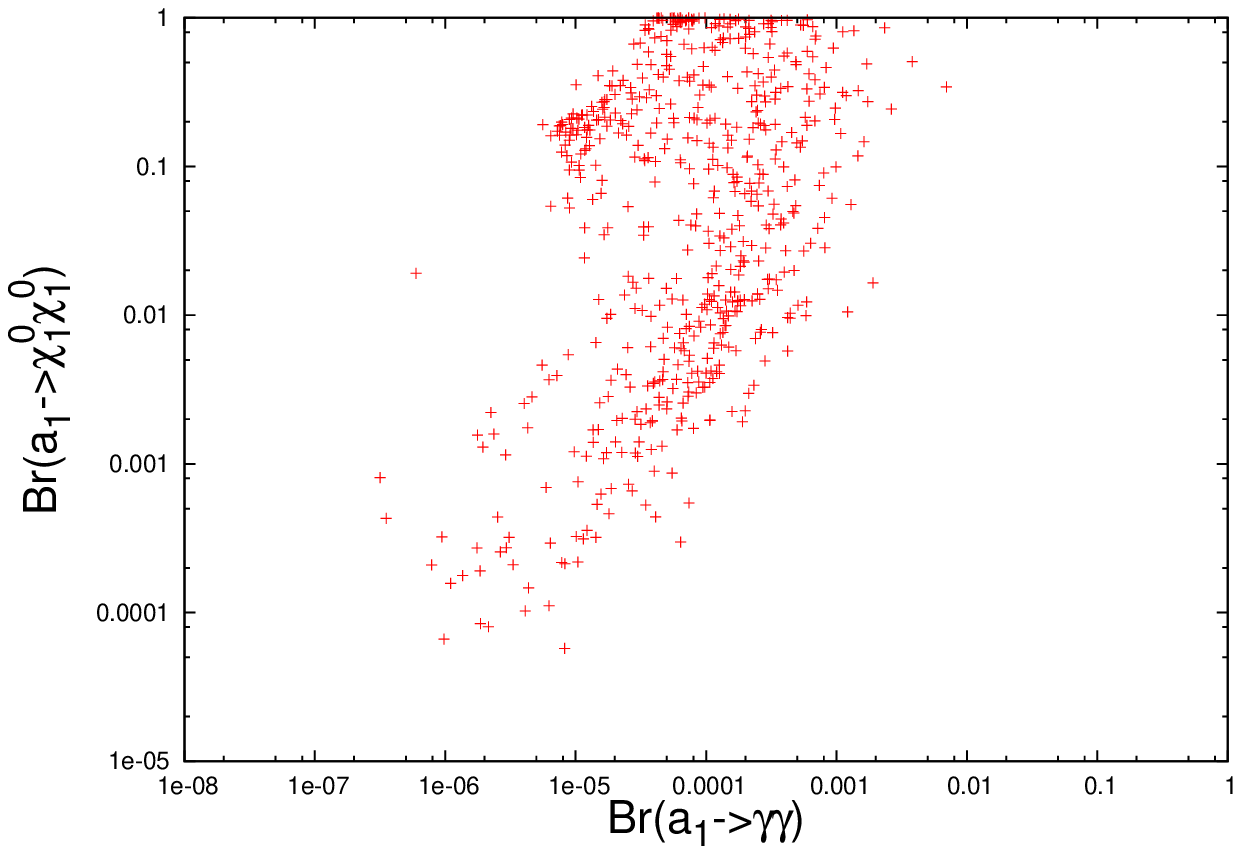}&\includegraphics[scale=0.60]{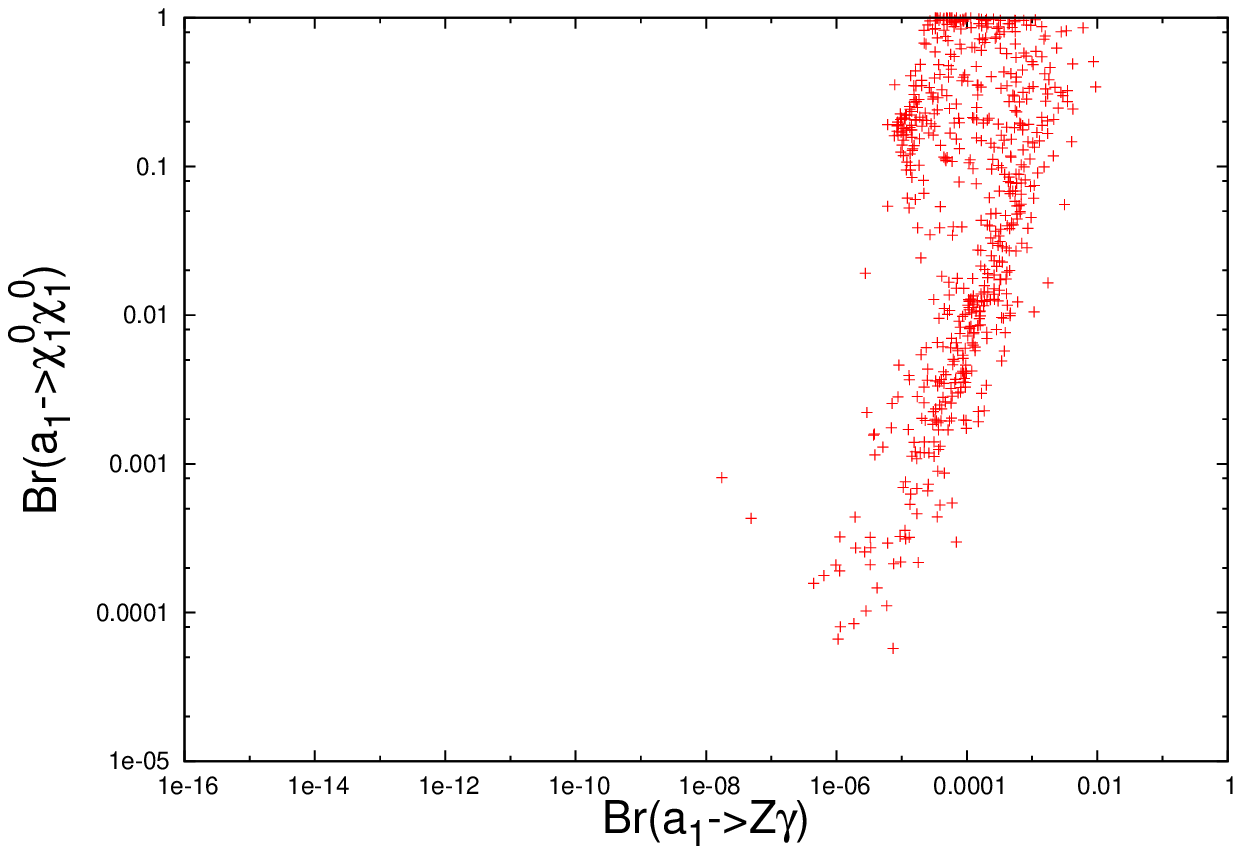}\\
  \includegraphics[scale=0.60]{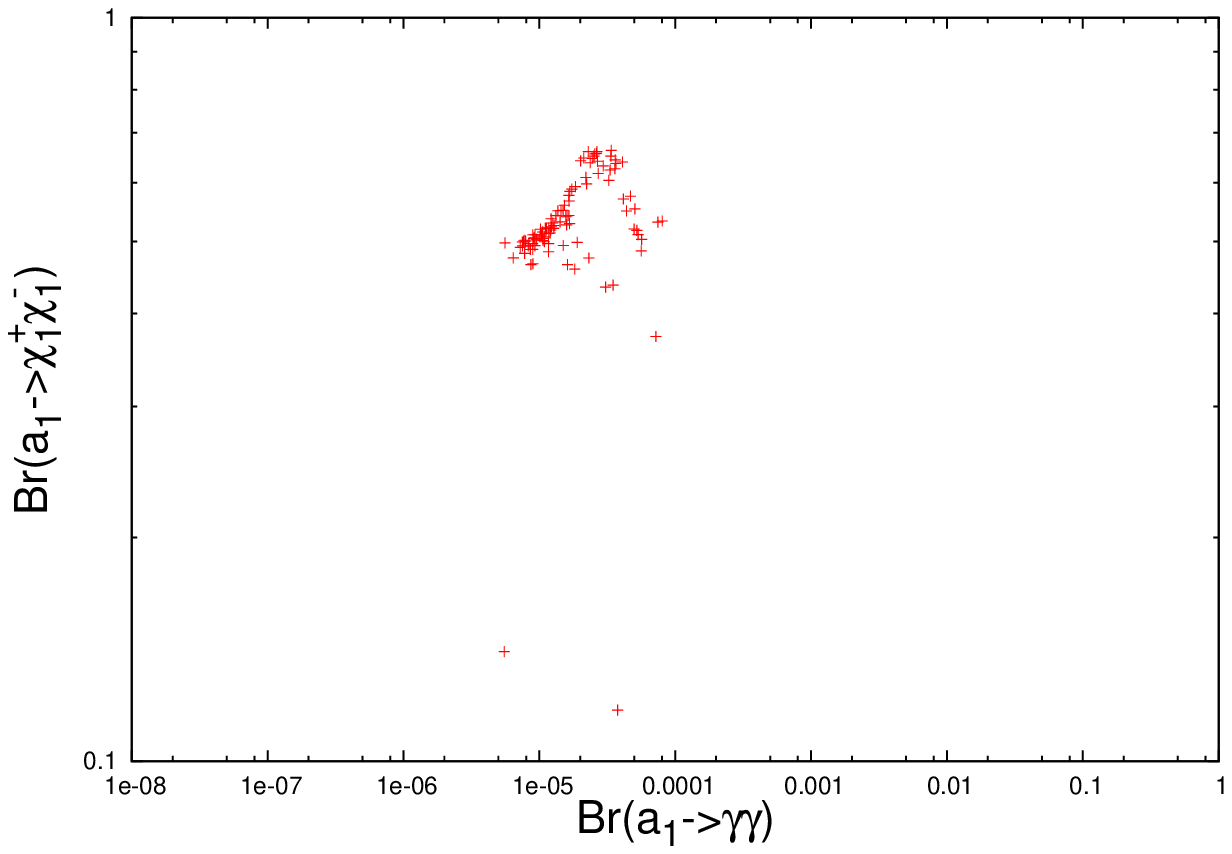}&\includegraphics[scale=0.60]{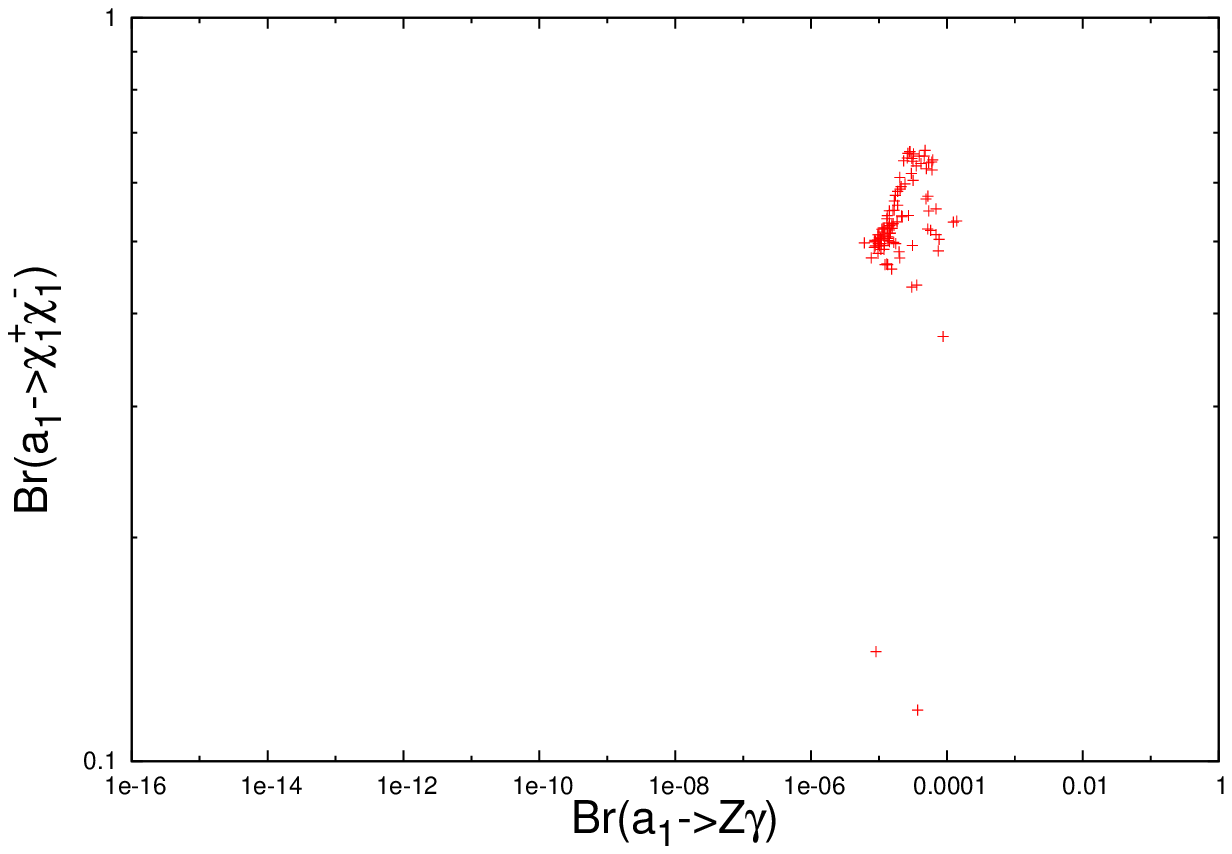}

 \end{tabular}
\label{fig:sigma-scan4}
\caption{The branching fractions ${\rm Br}(a_1\to \chi^0_1 \chi^0_1)$ (top) and ${\rm Br}(a_1\to \chi^+_1 \chi^-_1 )$ (bottom)
plotted against the branching fractions ${~\rm Br}(a_1\to \gamma\gamma)$ (left) 
and ${~\rm Br}(a_1\to Z\gamma)$ (right).}

\end{figure}

\end{document}